%% file: template2025.tex
\documentclass[9pt,conference]{IEEEtran}
\usepackage{dcase2025}

\usepackage{dcase2025,amsmath,graphicx,url,times,booktabs, tabularx
, bm, amssymb}

\title{AudioBERTScore: Objective Evaluation of Environmental Sound Synthesis Based on Similarity of Audio embedding Sequences}

\name{Minoru Kishi$^{1}$, Ryosuke Sakai$^{1}$, Shinnosuke Takamichi$^{1,2}$, Yusuke Kanamori$^{2}$, Yuki Okamoto$^{2}$}
\address{$^1$ Keio University, Japan. \{minorukishi1091, shinnosuke\_takamichi\}@keio.jp\\
    $^2$ The University of Tokyo, Japan. 
 }

\begin{document}

% \makeatletter
% \def\IEEEbibitemsep{0pt plus .3pt}
% \makeatother

%\setlength{\abovedisplayskip}{3pt} % 式の上部のマージン
%\setlength{\belowdisplayskip}{3pt} % 式の下部のマージン
\maketitle

\begin{abstract}
We propose a novel objective evaluation metric for synthesized audio in text-to-audio (TTA), aiming to improve the performance of TTA models.
In TTA, subjective evaluation of the synthesized sound is an important, but its implementation requires monetary costs. Therefore, objective evaluation such as mel-cepstral distortion are used, but the correlation between these objective metrics and subjective evaluation values is weak.
Our proposed objective evaluation metric, AudioBERTScore, calculates the similarity between embedding of the synthesized and reference sounds. The method is based not only on the max-norm used in conventional BERTScore but also on the $p$-norm to reflect the non-local nature of environmental sounds.
Experimental results show that scores obtained by the proposed method have a higher correlation with subjective evaluation values than conventional metrics.
\end{abstract}


\input{sections/1.introduction}

\input{sections/2.related-work}

\input{sections/3.proposed}

\input{sections/4.experiments}

\input{sections/5.conclusion}

{\footnotesize
\textbf{Acknowledgement:} This research was supported by JSPS KAKENHI 23K18474 and 25K21221, JST FOREST JP23KJ0828, and Moonshot R\&D Grant Number JPMJPS2011.
}

% -------------------------------------------------------------------------
% Either list references using the bibliography style file IEEEtran.bst

\clearpage
% The \IEEEtriggeratref{XX} command can be used to move to the next column before the XX-th reference
% to balance the two columns of the reference section
% \IEEEtriggeratref{XX}
\bibliographystyle{IEEEtran}
\bibliography{bib}

\end{document}

%% file: sections/1.introduction.tex
% \vspace{-2mm}
\section{Introduction}
% \vspace{-2mm}

Text-to-audio (TTA) models are deep learning models that generate environmental sounds from text inputs such as "a small dog is barking."
Synthesized audio are used for media content creation~\cite{tta-contents} and expressing characters’emotions.
The performance of TTA models is evaluated by the synthesized audio, and subjective evaluation is considered the most important~\cite{subscore-important}.
In fact, the final evaluation of TTA in the DCASE 2024 Challenge Task7\footnote{\url{https://dcase.community/challenge2024/task-sound-scene-synthesis}} is based on subjective evaluation.

Subjective evaluation of synthesized audio is usually conducted in terms of overall quality (OVL)~\cite{ovl} and relevance to the input text (REL)~\cite{subscore-important}, but it requires considerable time and financial costs.
Therefore, objective metrics such as mel cepstral distortion (MCD)~\cite{MCD} have been proposed.
It is essential to examine their correlation with subjective evaluations~\cite{subscore-important}; however, existing objective metrics show low correlation with subjective evaluation scores~\cite{human-clapscore}.

\begin{figure}[t]
\centering
\includegraphics[width=0.85\columnwidth]{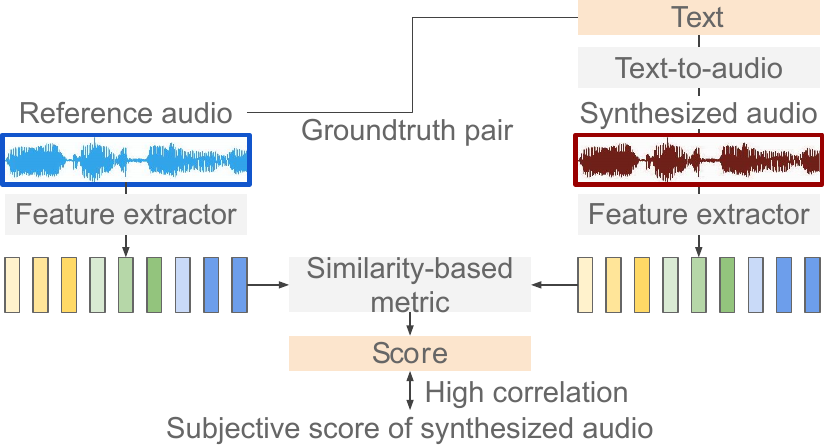} % fig/abs-system-outline.pdf
% \vspace{-3mm}
\caption{Overview of our AudioBERTScore.}
\label{fig:system-outline}
% \vspace{-4mm}
\end{figure}

In this study, we propose a new objective evaluation metric, AudioBERTScore, as shown in Figure\ref{fig:system-outline}.
This method uses both synthesized and reference audio. Embedding sequences are extracted from each audio using a audio foundation models like a ATST-Frame~\cite{atstframe}, and the evaluation score is estimated based on similarity between the sequences.
Experimental results show that among objective metrics using both synthesized and reference audio, score of the proposed method correlates most strongly with subjective evaluation scores.
The code is open-sourced at the project page\footnote{\url{https://github.com/lourson1091/audiobertscore}}.

%% file: sections/2.related-work.tex
% \vspace{-2mm}
\section{Related work}
% \vspace{-2mm}

\subsection{Evaluation metrics for synthesized audio of TTA}
% \vspace{-2mm}
\label{chap:tta-metric}
Subjective evaluation include OVL and REL~\cite{ovl-rel}. The former rates the quality of the synthesized audio, while the latter assesses the correspondence between the input text and the synthesized audio. Subjective evaluation is a useful means of quantifying perceived quality but requires monetary costs.

Therefore, objective evaluation methods that can predict scores correlated with subjective scores have been studied. In this paper, we classify the methods according to the following perspectives:

% \vspace{-1mm}
\begin{itemize} %\itemsep 0mm \leftskip -1mm
\item \textbf{Supervised training: }A direct approach is to train a model to predict scores from the synthesized audio, with supervised training on text, synthesized audio and subjective scores. We distinguish methods by the existence of such data and training.

\item \textbf{Reference: }Whether reference data is used in score estimation, and what type of reference is used e.g., text-reference-aware, audio-reference-aware, or reference-free.

\item \textbf{Pretrained model: }Whether a pretrained model is used in the estimation. If pretrained models are used, methods are further classified by the type of data used in pretraining.
\end{itemize}
% \vspace{-1mm}

We classify existing methods based on these criteria. They are shown in Table~\ref{tab:metric-classification} and as follows.

\input{tables/metric-classification}

\begin{itemize} %\itemsep 0mm \leftskip -1mm
\item \textbf{MCD, WARP-Q: }MCD~\cite{dtw} and WARP-Q~\cite{sdtw} compute distance between signal processing features of synthesized and reference audio. Since these scores can be calculated solely from the reference audio, they have the advantage of language independency of the input text. However, they have low correlation with subjective scores.

\item \textbf{RELATE benchmark: }This method~\cite{relate} was proposed as a supervised training model trained on sets of synthesized or natural audio, text and REL scores. Supervised training can achieve high correlation but requires large-scale datasets of subjective scores.

\item \textbf{PAM: }This method~\cite{pam} uses contrastive learning of text and audio to estimate quality scores from text prompts referring to audio quality and the synthesized audio. PAM does not require reference, but since it depends on pretraining with paired text-audio data, it can only be used for languages where such data exists.
\end{itemize}
% \vspace{-1mm}
Our AudioBERTScore uses reference audio without supervised training, like MCD and WARP-Q. While maintaining the strength of language independence, it aims for high correlation with subjective scores by improving feature extraction and similarity calculation using foundation models.

% \vspace{-2mm}
\subsection{SpeechBERTScore}
% \vspace{-2mm}
SpeechBERTScore~\cite{speechbertscore-en}, which is originated from BERTScore~\cite{bert-score} in natural language processing, was proposed to automatically evaluate synthesized speech in text-to-speech.
The original BERTScore calculates a similarity score between embedding vector sequences obtained from generated and reference sentences. It achieves high correlation with subjective scores by using BERT~\cite{bert}. SpeechBERTScore successfully applies this framework to synthesized speech by replacing BERT with speech-specific foundation models~\cite{hubert}.

Our AudioBERTScore also follows this trend. Specifically, it uses audio foundation models for the extraction. Furthermore, we design similarity calculation for environmental sound.
% BERTScore calculates a series of contextual embedding vectors from both the generated and reference sentences and evaluates similarity between these series. It achieves high correlation with subjective evaluations by using the Bidirectional Encoder Representations from Transformers (BERT)~\cite{bert} self-supervised learning model for embedding, and by computing similarity with forced alignment of the sequences.

% SpeechBERTScore successfully applies this framework to synthesized speech by replacing the self-supervised model with audio-specific models (e.g., HuBERT\cite{hubert}, WavLM\cite{wavlm}).

%% file: tables/metric-classification.tex
{
\tabcolsep = 3pt
\begin{table}[t]
%\vspace{-1mm}
\caption{Classification of objective metric for synthesized audio}
\label{tab:metric-classification}
\centering
%\scriptsize
\begin{tabular}{p{0.3\linewidth}|ccc}
Method                      & Training  & Reference     & Pretrained model\\ \hline
MCD              & No            & Yes (audio)   & No\\
WARP-Q           & No            & Yes (audio)   & No\\
FAD              & No            & Yes (audio)   & Yes (audio)\\
\textbf{AudioBERTScore (ours)}   & No            & Yes (audio)   & Yes (audio)\\ \hline
RELATE bench   & Yes           & Yes (text)    & Yes (text \& audio)\\
CLAPScore    & No            & Yes (text)    & Yes (text-audio)\\
PAM              & No            & No            & Yes (text-audio)
\end{tabular}
%\vspace{-4mm}
\end{table}
}

%% file: sections/3.proposed.tex
% \vspace{-2mm}
\section{Proposed objective evaluation metric}
\label{sec:system-construction}
% \vspace{-2mm}
\subsection{Feature extraction and similarity matrix}
% \vspace{-2mm}
\label{chap:encode}
We first obtain the embedding sequences for both the synthesized and reference audio. 
Let the waveform of the synthesized audio be represented as
$\bm{s} = (s_t \in \mathbb{R} \mid t=1, \ldots, T_{\text{gen}})$, and
that of the reference audio as
$\bm{r} = (r_t \in \mathbb{R} \mid t=1, \ldots, T_{\text{ref}})$.
$T_{\text{gen}} \neq T_{\text{ref}}$ in general.

The embedding sequences extracted from $\mathit{S}$ and $\mathit{R}$ using a feature extractor are represented as:
$
\tilde{\mathit{S}} = (\tilde{\bm{s}}_n \in \mathbb{R}^D \mid n=1, \ldots, L_{\text{gen}}), \quad
\tilde{\mathit{R}} = (\tilde{\bm{r}}_n \in \mathbb{R}^D \mid n=1, \ldots, L_{\text{ref}})
$
These are computed as
$
\tilde{\mathit{S}} = \text{Encoder}(\bm{s}; \theta) 
,
\tilde{\mathit{R}}  = \text{Encoder}(\bm{r}; \theta).
$
$\theta$ represents the parameters of a pretrained feature extractor. 
$L_{\text{gen}}$ and $L_{\text{ref}}$ are determined by $T_{\text{gen}}$ and $T_{\text{ref}}$.

% The choice of feature extractor is arbitrary. For instance, when using a self-supervised model based on Transformers~\cite{transformer}, it is expected that shallow layers capture acoustic features while deeper layers capture contextual features.

Using $\tilde{\mathit{S}}$ and $\tilde{\mathit{R}}$, we compute similarities between each pair of embeddings and represent them in matrix form. The similarity matrix $\mathit{M} \in \mathbb{R}^{L_{\text{gen}} \times L_{\text{ref}}}$ is defined by cosine similarity for each element $(i,j)$ as:
\begin{align}
\mathit{M}_{ij} = \mathrm{sim}(\tilde{\bm{s}}_i, \tilde{\bm{r}}_j) =
\frac{ \tilde{\bm{s}}_i \cdot \tilde{\bm{r}}_j }
     { \left\lVert \tilde{\bm{s}}_i \right\rVert \cdot \left\lVert \tilde{\bm{r}}_j \right\rVert }
\end{align}

% \vspace{-2mm}
\subsection{Score calculation}
\begin{figure}[t]
    \centering
    \includegraphics[width=\linewidth]{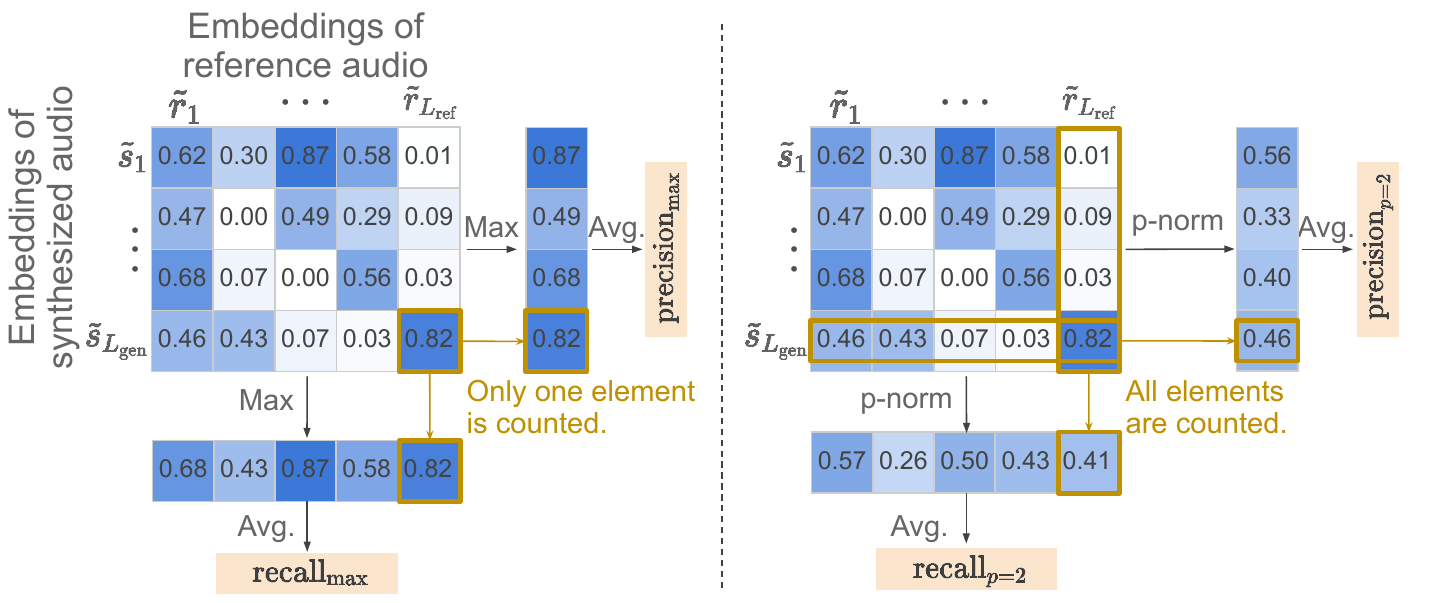}
    % \vspace{-5mm}
    \caption{Similarity computation in our method. Max-norm (left) assumes locality, and $p$-norm (right) reflects the non-locality.% of environmental sounds. %$\mathrm{precison}_{*}, \mathrm{recall}_{*}$ represent averages of aggregated row- or column-wise values of the similarity matrix. The max-based approach (left) selects maximum values, while the $p$-norm approach (right) uses the entire set of elements.
    }
    % \vspace{-4mm}
    \label{fig:metric-calculation}
\end{figure}

% \vspace{-2mm}
Scores are calculated based on the similarity matrix. We first apply a method based on max-norm, as used in BERTScore and SpeechBERTScore. Then, considering non-locality of environmental sounds, we propose a method based on the $p$-norm. Figure~\ref{fig:metric-calculation} shows the computation.

% \vspace{-1mm}
\subsubsection{Computation based on maximum scores}
% \vspace{-1mm}
\label{chap:normal-calc}
We compute precision, recall, and F1 score from the similarity matrix. Precision is the average of the maximum similarity for each frame in the synthesized embeddings, representing how well the synthesized audio covers the reference. Recall is computed by swapping synthesized and reference audio, indicating how well the reference is covered by the synthesized one. The harmonic mean of these scores gives the F1 score. These are calculated as
\begin{align}
\label{eq:precision-max}
\mathrm{precision}_{\mathrm{max}} &= \frac{1}{L_{\mathrm{gen}}} \sum\limits_{i=1}^{L_{\mathrm{gen}}} \max_{j = 1, \ldots, L_{\mathrm{ref}}} \mathit{M}_{ij}\\
\label{eq:recall-max}
\mathrm{recall}_{\mathrm{max}} &= \frac{1}{L_{\mathrm{ref}}} \sum\limits_{j=1}^{L_{\mathrm{ref}}} \max_{i = 1, \ldots, L_{\mathrm{gen}}} \mathit{M}_{ij} \\
\label{eq:f1-max}
\mathrm{F1}_{\mathrm{max}} &= 2 \times 
    \frac{\text{precision}_{\mathrm{max}} \times \text{recall}_{\mathrm{max}}}
    {\text{precision}_{\mathrm{max}} + \text{recall}_{\mathrm{max}}} 
\end{align}

These scoring methods use the $\infty$-norm (max-norm), which assumes high-similarity vectors are temporally localized. This assumption generally holds for natural language and speech, where phrases and segmental features are temporally bounded.

However, this assumption may not hold for environmental sounds. For example, a gunshot sound is a localized, instantaneous sound where embeddings cluster temporally, making locality assumptions valid. In contrast, unstructured, continuous sounds like a babbling brook may have embeddings spread across time, violating this assumption. Therefore, a scoring method that can capture such non-local characteristics is needed.

% \vspace{-1mm}
\subsubsection{Computation based on $p$-norm}
% \vspace{-1mm}
Replacing the max-norm with the $p$-norm, we define the following scores:
\begin{align}
    \mathrm{precision}_{p}  &= \frac{1}{L_{\mathrm{gen}}} \sum\limits_{i=1}^{L_{\mathrm{gen}}}
    \left(\frac{1}{L_{\mathrm{ref}}}  \sum\limits_{j=1}^{L_{\mathrm{ref}}} M_{ij}^p \right)^{1/p} \\
    \mathrm{recall}_{p}     &= \frac{1}{L_{\mathrm{ref}}} \sum\limits_{j=1}^{L_{\mathrm{ref}}}
    \left(\frac{1}{L_{\mathrm{gen}}}  \sum\limits_{i=1}^{L_{\mathrm{gen}}} M_{ij}^p \right)^{1/p} 
\end{align}

As shown in the right part of Figure~\ref{fig:metric-calculation}, these are computed using the $p$-norm. When $p = 1$, the metrics are simple averages and measure overall (non-local) similarity across time. As $p$ increases, more weight is placed on local peaks, capturing locality. When $p \to \infty$, the score is equivalent to the max-norm.

To balance local and non-local similarity, we introduce the following interpolation between max-based and $p$-norm-based scores:
\begin{align}
    \label{eq:precision_plam}
    \mathrm{precision}_{\lambda, p} 
        &= \lambda \cdot \mathrm{precision}_{\textrm{max}} + (1 - \lambda)\cdot \mathrm{precision}_{p}  \\
    \label{eq:recall_plam}
    \mathrm{recall}_{\lambda, p}    
        &= \lambda \cdot \mathrm{recall}_{\textrm{max}} + (1 - \lambda)\cdot \mathrm{recall}_{p}  \\
    \label{eq:f1_plam}
    \mathrm{F1}_{\lambda, p}
        &= 2 \times 
        \frac{\text{precision}_{\lambda, p} \times \text{recall}_{\lambda, p}}{\text{precision}_{\lambda, p} + \text{recall}_{\lambda, p}} 
\end{align}
$\lambda \in [0, 1]$ is a hyperparameter for interpolation. When $\lambda = 1$ or $p \to \infty$, the score is equivalent to the max-norm.

%% file: sections/4.experiments.tex
% \vspace{-2mm}
\section{Experimental evaluation}
% \vspace{-2mm}

\label{chap:result}
To evaluate our method, we computed the correlation between the automatic scores calculated by our method and the subjective scores.

% \vspace{-2mm}
\subsection{Experimental conditions}
% \vspace{-2mm}
\label{ssec:exp:sds:config}

\textbf{Dataset.} \label{ssec:exp:sds:method}
We used the PAM test set~\cite{pam}, a dataset consisting of synthesized and reference audio, English text, and subjective scores. This set includes 100 pairs of natural audio (reference) randomly extracted from AudioCaps~\cite{audiocaps} and their captions, along with 400 synthesized audio by MelDiffusion, AudioLDM 2\footnote{\url{https://github.com/haoheliu/AudioLDM2}}~\cite{audioldm2}, AudioLDM-Large\footnote{\url{https://github.com/haoheliu/AudioLDM}}~\cite{audioldm}, and AudioGen-base\footnote{\url{https://github.com/facebookresearch/audiocraft}}~\cite{audiogen}. Each sample of both reference and synthesized audio is annotated with 5-point MOS scores for OVL and REL. Each MOS score is the average of scores given by 10 different raters. We excluded from 17 reference audio samples with REL subjective scores less than $3.5$ and the corresponding $17 \times 4$ synthesized audio samples. This exclusion was made to ensure the proposed method accurately estimates scores correlated to the REL subjective scores, which assume a strong relation between the reference audio and text. The duration of each sample is 5 seconds for synthesized audio and 10 seconds for reference audio. All audio samples were downsampled at $16$~kHz.

% The four synthesis models used were: MelDiffusion, AudioLDM2\footnote{\url{https://github.com/haoheliu/AudioLDM2}}, AudioLDM-Large\footnote{\url{https://github.com/haoheliu/AudioLDM}}\cite{audioldm}, and AudioGen-base\footnote{\url{https://github.com/facebookresearch/audiocraft}}\cite{audiogen}. For details about the models, refer to PAM~\cite{pam}. Sampling rates were 44.1~kHz for reference audio, and 22.05~kHz or 16~kHz for synthesized audio. All were downsampled to 16~kHz before being input to the feature extractors described later.

\textbf{Feature extractors.}
We used the following three pretrained models as feature extractors for the proposed method.
% \vspace{-1mm}
\begin{itemize} %\itemsep 0mm \leftskip -1mm
    \item \textbf{BYOL-A~\cite{BYOL-A2}\footnote{\url{https://github.com/nttcslab/byol-a/blob/master/v2/AudioNTT2022-BYOLA-64x96d2048.pth}}}: A model based on convolutional neural networks (CNN)~\cite{cnn}. We used the latest v2. Frame-level embeddings (``local'') , its channel-flattened feature (``global''), and their concatenation along the feature dimension (``local+global'') were used. 
    \item \textbf{ATST-Frame~\cite{atstframe}\footnote{\url{https://github.com/Audio-WestlakeU/audiossl/blob/main/audiossl/methods/ATST-Frame/README.md}}}: A 13-layer Transformer~\cite{transformer}-based model. Features from each of the 1st--13th layers were used. ATST-Frame-base model was used.
    \item \textbf{AST~\cite{AST}\footnote{\url{https://github.com/YuanGongND/ast/blob/master/pretrained_models/README.md}}}: A 13-layer Transformer~\cite{transformer}-based model finetuned on environmental sound classification. Features from each of the 1st through 13th layers were used. We used the fine-tuned model with Full AudioSet, 10 tstride, 10 fstride, and weight averaging (0.459 mAP).
\end{itemize}
% \vspace{-1mm}

\textbf{Comparison methods.}
As comparison methods under the same conditions (Table~\ref{tab:metric-classification}), we used MCD~\cite{MCD} and WARP-Q~\cite{warp-q}, which do not require training and use reference audio. Although FAD~\cite{fad} also falls under the same condition, it is excluded from comparison since it is calculated over multiple samples, unlike the proposed method, MCD, and WARP-Q which are computed per sample. Additionally, as methods under different conditions, we also included CLAPScore~\cite{clapscore} and PAM~\cite{pam}.

\textbf{Evaluation method.}
For each of the OVL and REL subjective scores, we computed the linear correlation coefficient (LCC) and Spearman’s rank correlation coefficient (SRCC) with the objective evaluation metrics. These coefficients were computed across all samples in the test set.

% \vspace{-2mm}
\subsection{Results}

% \vspace{-2mm}
\subsubsection{Feature extractors with precision, recall, and F1}
\label{chap:result-encoder}
% \vspace{-1mm}

\label{chap:result-layer}
\textbf{Comparison of feature extractors.}
\begin{figure}[t]
\centering
\includegraphics[width=0.98\linewidth]{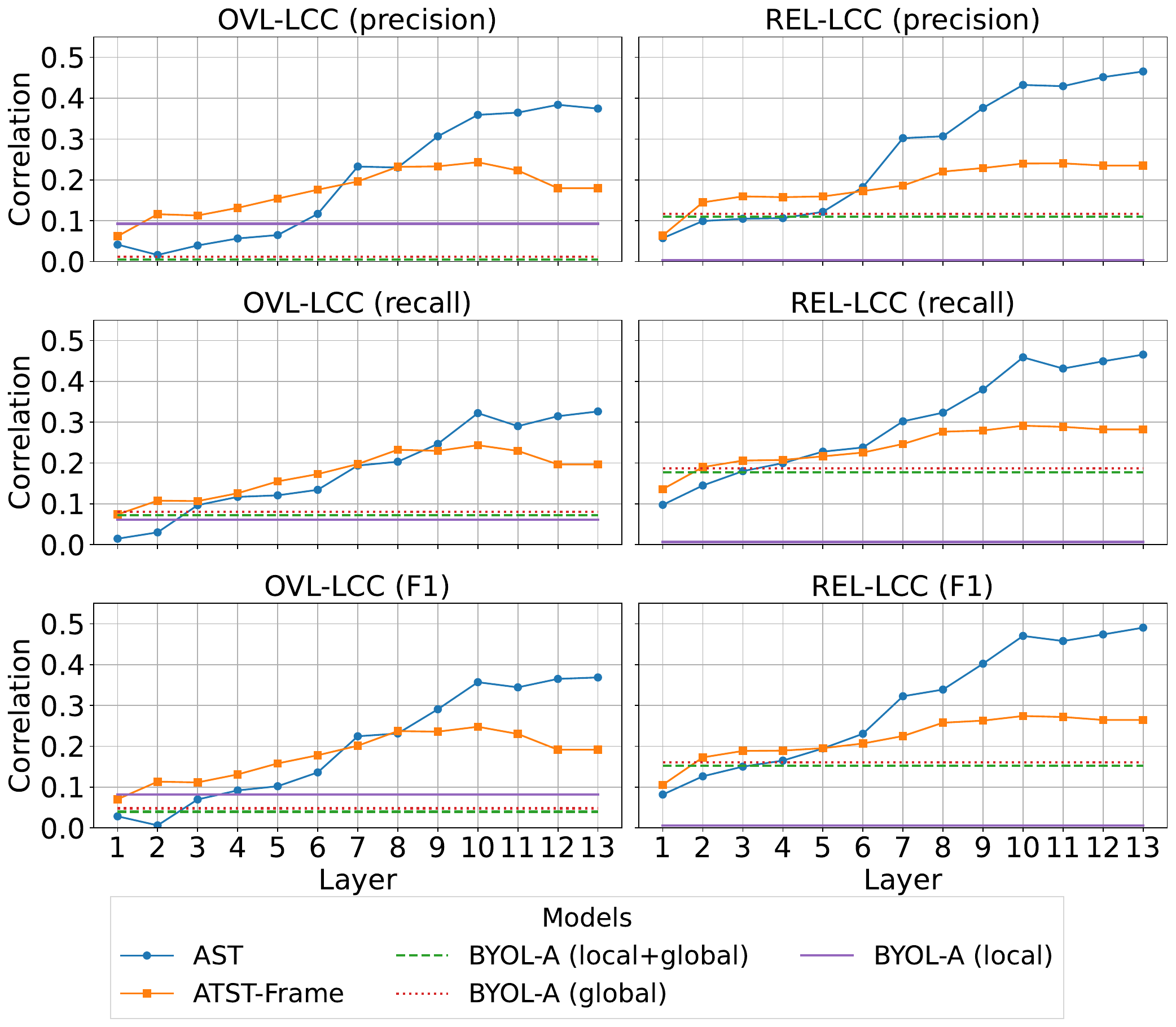} 
% \vspace{-3mm}
\caption{Correlation under several settings of feature extractors and similarity computation.}
\label{fig:layer-graph}
% \vspace{-4mm}
\end{figure}

\input{tables/result-encoder}
We compared the correlation between scores calculated from each layer of the feature extractors and the subjective evaluation scores. The scores were computed using the max-based method (Equations~(\ref{eq:precision-max})--(\ref{eq:f1-max})). Results are shown in Figure~\ref{fig:layer-graph}\footnote{Due to the page limit, we only show results of LCC, but those of SRCC are the same trends.}. Both AST and ATST-Frame showed higher correlation in later layers, indicating that later layers capture contextual information relevant for environmental sounds, which benefits AudioBERTScore. For BYOL-A, the OVL score is higher for the local embedding, while the REL score is higher for global or local+global, likely because local captures acoustic features near the input layer, while global retains contextual information.

\begin{figure}[t]
\centering
\includegraphics[width=0.50\linewidth]{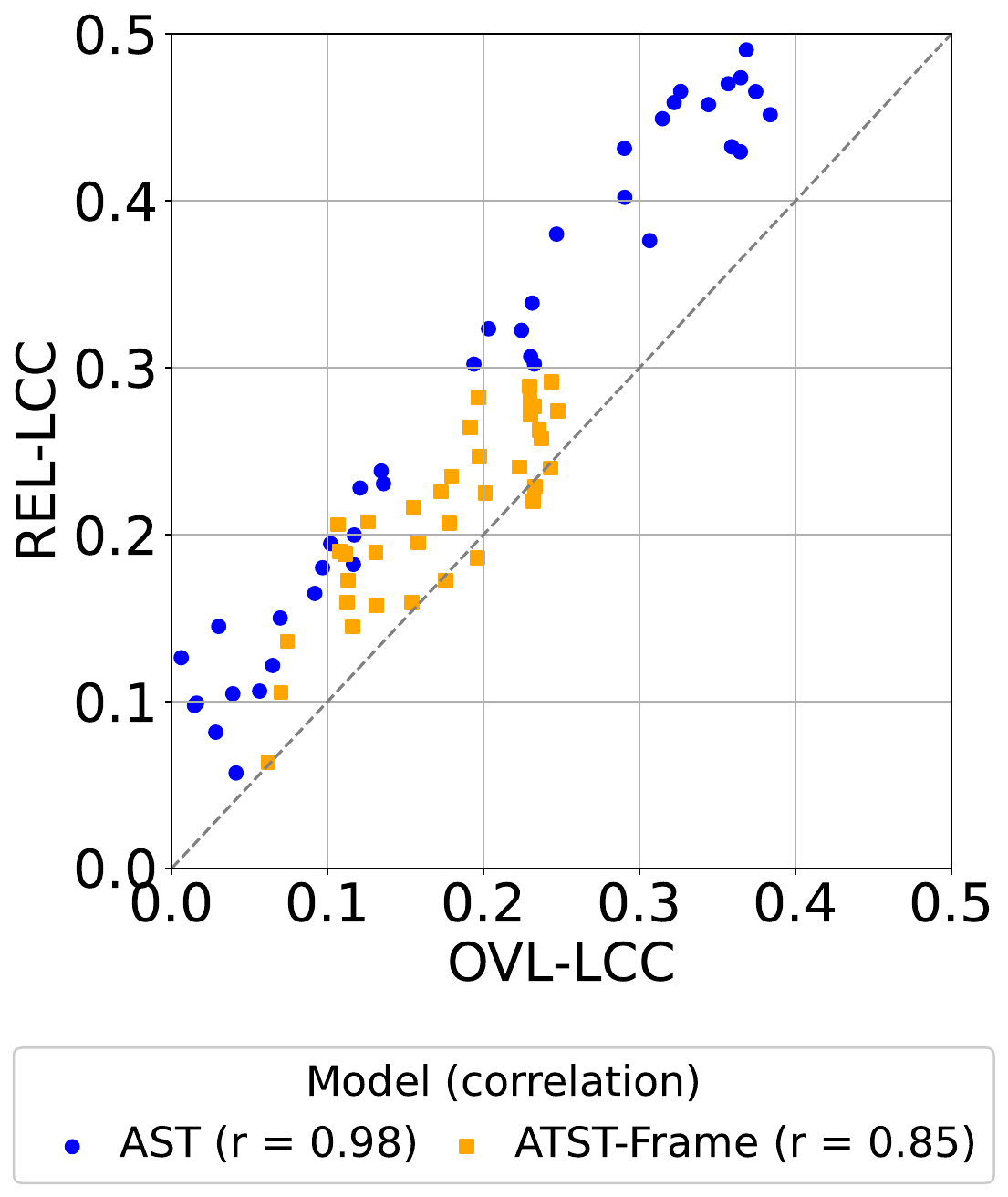} 
% \vspace{-3mm}
\caption{Scatter plot of OVL-LCC vs REL-LCC.Precision, recall, and F1 metrics across all 13 layers are included.  The $r$ value in the legend is the Pearson correlation coefficient.}
% \vspace{-4mm}
\label{fig:ovl-rel-graph}
\end{figure}

\textbf{Correlation between correlation scores of OVL-REL.}
Focusing on correlation scores of OVL and REL for AST and ATST-Frame in Figure~\ref{fig:layer-graph}, there is a trend that when correlation with OVL is high, correlation with REL is also high. To examine this, we show the correlation between OVL-LCC and REL-LCC in Figure~\ref{fig:ovl-rel-graph}. Both models show high correlation between these two metrics. Also, most data points lie above the diagonal, suggesting REL-LCC tends to be higher than OVL-LCC, likely because AudioBERTScore is well-designed to capture contextual relevance.

\textbf{Comparison of precision, recall, and F1.}
Figure~\ref{fig:layer-graph} indicates that both AST and ATST-Frame show a tendency for higher recall than precision. In AST, precision rapidly increases around layers 6–7, and recall increases sharply around layers 8–10. Further investigation is needed to explain these patterns.

\textbf{Comparison of best configurations.}
Table~\ref{tab:result-encoder} shows the results for the best settings for each feature extractor, chosen based on average high correlation scores in Figure~\ref{fig:layer-graph}.
%: AST uses $\mathrm{F1}_{\mathrm{max}}$ from the 13th layer, ATST-Frame uses $\mathrm{recall}_{\mathrm{max}}$ from the 10th layer, and BYOL-A uses $\mathrm{recall}_{\mathrm{max}}$ from the global layer. 
The Transformer-based models AST and ATST-Frame significantly outperform the CNN-based BYOL-A, indicating the effectiveness of contextual information extraction with Transformers~\cite{transformer-than-cnn}. Furthermore, AST’s superior performance suggests that fine-tuning for environmental sound classification contributed to the improvement.

% \vspace{-1mm}
\subsubsection{Max-norm vs. $p$-norm}
\label{chap:result-p}
% \vspace{-1mm}

\begin{figure}[t]
\centering
\includegraphics[width=0.98\linewidth]{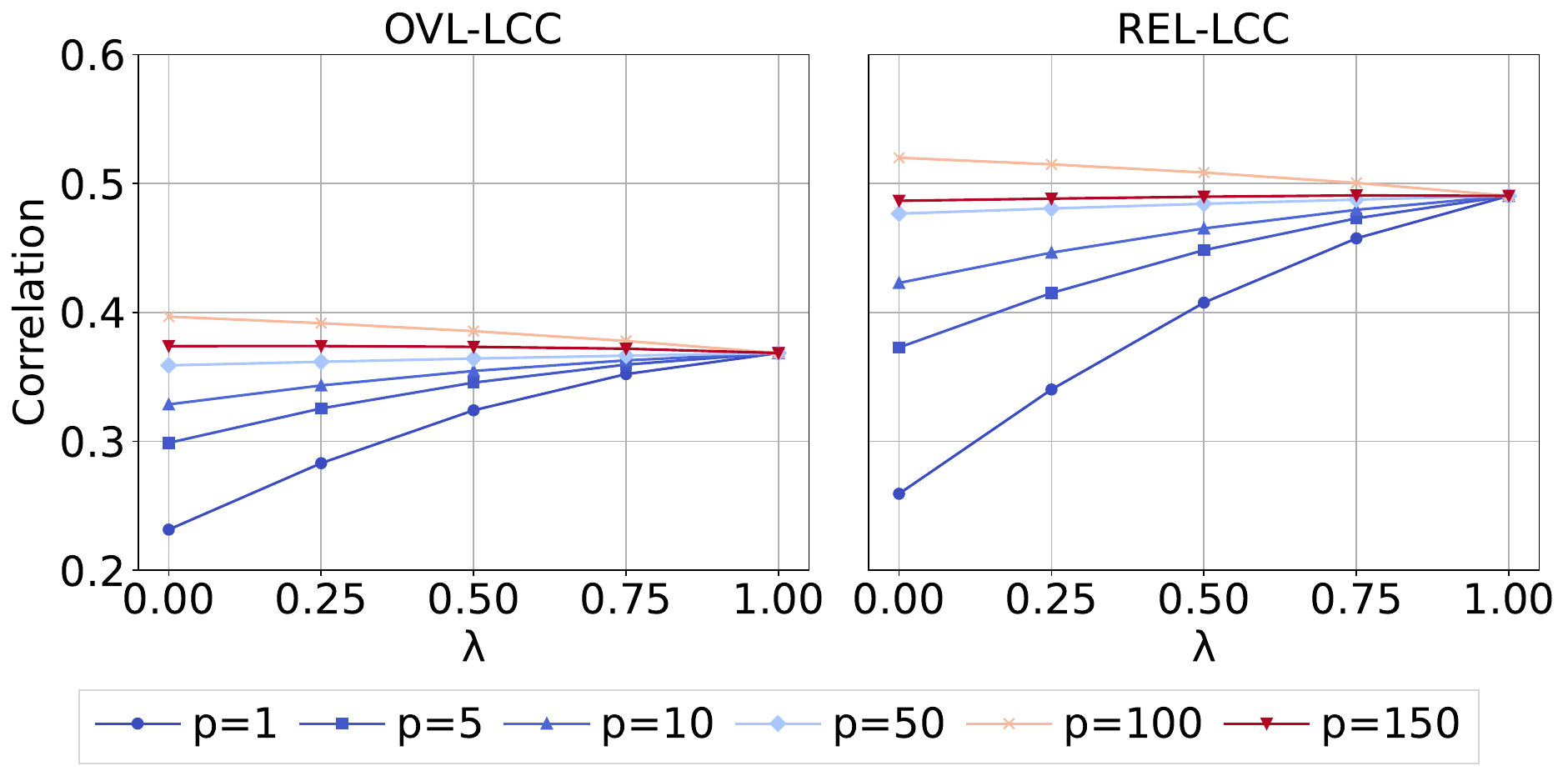} 
% \vspace{-3mm}
\caption{Correlation using $p$-norm-based calculation with various $p$ and $\lambda$.}
% \vspace{-4mm}
\label{fig:p-graph}
\end{figure}

\begin{figure}[t]
\centering
\includegraphics[width=0.98\linewidth]{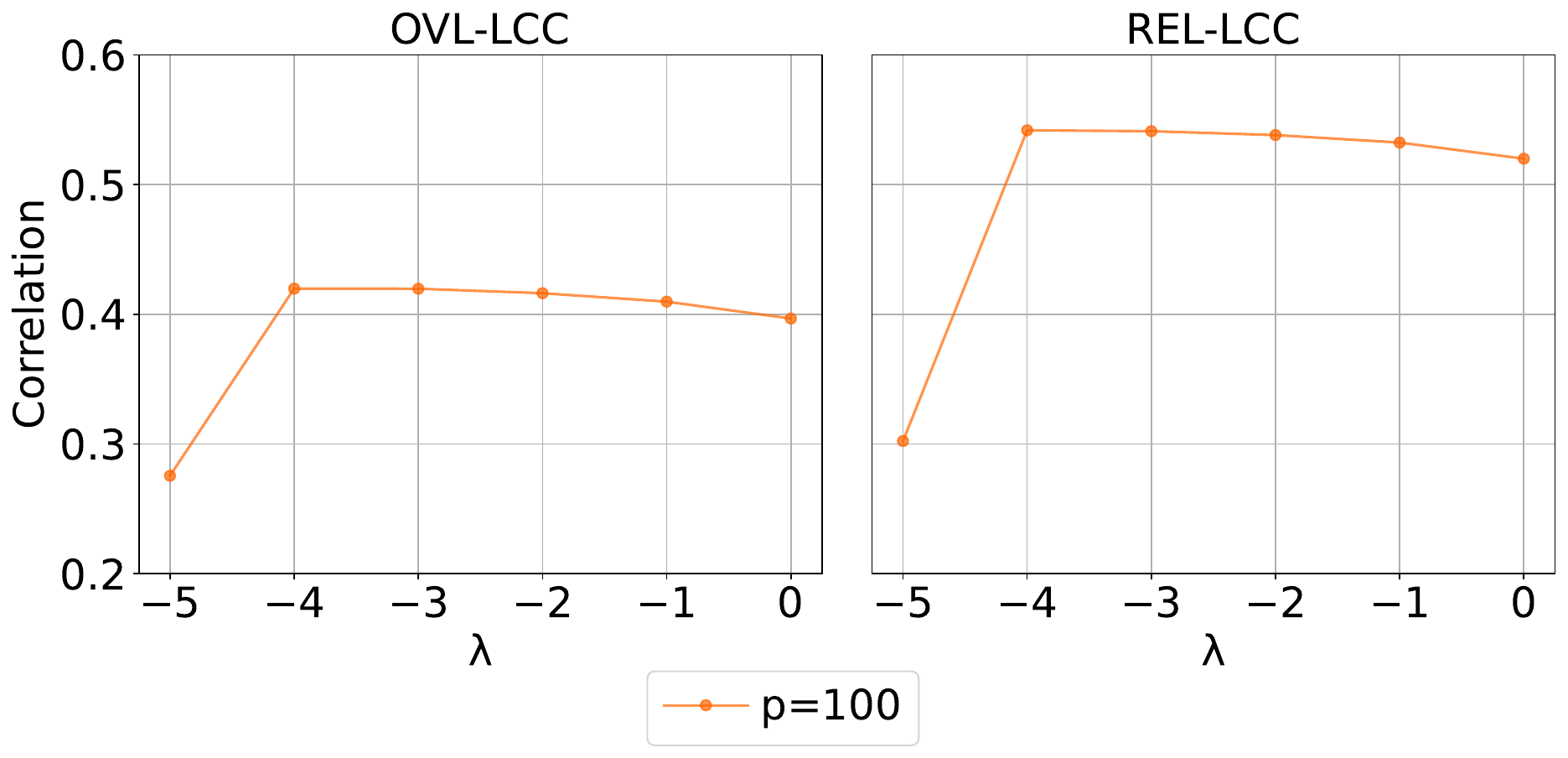} 
% \vspace{-3mm}
\caption{Correlation using $p=100$ and negative values of $\lambda$.}
% \vspace{-4mm}
\label{fig:p-minus-graph}
\end{figure}

\textbf{Effect of $p$ and $\lambda$.}
We investigated the performance of the $p$-norm based score (Equations~(\ref{eq:precision_plam})--(\ref{eq:f1_plam})) with various values of $p$ and $\lambda$. Results using the 13th layer F1 score of AST (the best performer in Section~\ref{chap:result-layer}) are shown in Figure~\ref{fig:p-graph}. Performance peaked at $p=100$, especially at $\lambda=0$. Interestingly, $p=106$ slightly outperformed $p=100$.

\textbf{Investigation of negative $\lambda$.}
From the slopes in Figure~\ref{fig:p-graph}, correlation increases as $\lambda$ decreases, even continuing beyond $\lambda=0$. Thus, we explored negative $\lambda$ values to assess potential further improvements. Figure~\ref{fig:p-minus-graph} shows correlation improves until $\lambda=-4$, then drops at $\lambda=-5$. We hypothesize that both localized and non-localized similarity contribute positively when the $p$-norm weight increases and maximum-based score grows in magnitude.

% \vspace{-1mm}
\subsubsection{Comparison with other objective metrics}
% \vspace{-1mm}
We compared the performance of the proposed method with existing objective metrics. Results are shown in Table~\ref{tab:result-metric}, using the best-performing configurations found in Sections~\ref{chap:result-encoder} and \ref{chap:result-p}, with $p=106$ and $\lambda=-3.5$ as derived from Figure~\ref{fig:p-minus-graph}. While not direct competitors, PAM and CLAPScore are included for reference.

The proposed method significantly outperformed MCD and WARP-Q in both REL and OVL correlations, showing its usefulness as a training-free, reference-based evaluation metric. Comparing internal variants, $\mathrm{F1}_{\mathrm{max}}$, $p$-norm $\mathrm{F1}_{\lambda=0, p=106}$, and extrapolated norm $\mathrm{F1}_{\lambda=-3.5, p=106}$ all contributed to performance improvements.

Lastly, we compare the proposed method with PAM and CLAPScore. The proposed method shows the highest correlation with REL, supporting its contribution. While PAM slightly outperforms in OVL, this will be addressed in future work.

\input{tables/result-metric}

%% file: tables/result-encoder.tex
{\tabcolsep 1.5mm
\begin{table}[t]
    %\vspace{-1mm}
  \centering
  \caption{Comparison for each feature extractor.The score calculation based on the maximum value and the features extracted use the best settings.}
  %\scriptsize
  \label{tab:result-encoder}
  \begin{tabular}{l|cc|cc}
                & \multicolumn{2}{c|}{OVL} & \multicolumn{2}{c}{REL} \\
                & LCC           & SRCC          & LCC           & SRCC \\ \hline
    AST, $13$th layer, $\mathrm{F1}_{\mathrm{max}}$         
                & $\bm{0.368}$  & $\bm{0.397}$  & $\bm{0.490}$  & $\bm{0.515}$ \\
    ATST-Frame, $10$th layer, $\mathrm{recall}_{\mathrm{max}}$  
                & $0.248$       & $0.308$       & $0.291$       & $0.343$ \\
    BYOL-A, global feature, $\mathrm{recall}_{\mathrm{max}}$      
                & $0.080$       & $0.077$       & $0.187$       & $0.196$
  \end{tabular}
  %\vspace{-4mm}
\end{table}
}

%% file: tables/result-metric.tex
\begin{table}[t]
  \centering
  %\vspace{-1mm}
  \caption{Evaluation results for each objective metric. Higher values indicate a stronger correlation with the subjective evaluation scores.}
  \label{tab:result-metric}
  %\scriptsize
  \begin{tabular}{l|cc|cc}
    & \multicolumn{2}{c|}{OVL} & \multicolumn{2}{c}{REL} \\ 
                    & LCC           & SRCC          & LCC           & SRCC \\
    \midrule
    \multicolumn{5}{l}{Compared metrics} \\
    \midrule
    MCD             & $0.004$       & $0.008$       & $0.029$       & $0.05$ \\
    WARP-Q          & $0.241$       & $0.228$       & $0.202$       & $0.213$ \\ 
    \midrule
    \multicolumn{5}{l}{Proposed AudioBERTScore (AST, $13$th layer)} \\
    \midrule
    $\mathrm{F1}_{\mathrm{max}}$
                    & $0.368$       & $0.397$       & $0.490$       & $0.515$ \\
    $\mathrm{F1}_{\lambda = 0, p = 106}$
                    & $0.401$         & $0.420$         & $0.523$         & $0.545$ \\
    $\mathrm{F1}_{\lambda = -3.5, p = 106}$
                    & $\bm{0.424}$  & $\bm{0.433}$  & $\bm{0.546}$  & $\bm{0.567}$ \\
    \midrule
    \multicolumn{5}{l}{Other metrics} \\
    \midrule
    %RELATE bench    & -             & -             & x.xxx         & x.xxx \\
    PAM             & 0.595         & 0.604         & 0.529         & 0.556 \\
    CLAPScore       & 0.337         & 0.323         & 0.487         & 0.475 \\
  \end{tabular}
  %\vspace{-4mm}
\end{table}

%% file: sections/5.conclusion.tex
% \vspace{-2mm}
\section{Conclusion}
% \vspace{-2mm}
In this paper, we proposed an objective evaluation metric for TTA based on the similarity between sequences of synthesized audio embedding.  
Evaluation results demonstrated that the method achieved the best performance among unsupervised, audio-reference metrics. Furthermore, it outperformed other metrics even under different evaluation conditions.

% As future work, we plan to explore improved similarity and score computation methods.  
% Additionally, during the course of this study, we identified a potential data leakage issue between the pretraining datasets of existing pretrained models and the commonly used evaluation sets.  
% To address this, we are considering the construction of a new evaluation dataset. Further details will be reported in a future publication.